\documentclass[12pt,onecolumn, peerreview,draftclsnofoot]{IEEEtran} 
\ifCLASSINFOpdf
   \usepackage[pdftex]{graphicx}
\else
   \usepackage[dvips]{graphicx}  
\fi
\graphicspath{{figures/}}
\usepackage[cmex10]{amsmath}
\usepackage{array}
\usepackage{eqparbox}
\usepackage{txfonts}
\usepackage{verbatim, amssymb}
\usepackage{mathrsfs}
\usepackage{bm}
\usepackage{lipsum}
\usepackage{cite}
\usepackage{algpseudocode}
\usepackage{algorithmicx}
\usepackage{algpseudocode}
\usepackage{algorithm}
\usepackage{textcomp}

\begin{document}
 
\title{Application of Asynchronous Weak Commitment Search in Autonomous Quality of Service Provision in Cognitive Radio Networks}

\author{Shabnam~Sodagari, shabnam@ieee.org}
\maketitle

\begin{abstract}
This article presents a distributed solution to autonomous quality of service provision in cognitive radio networks. Specifically, cognitive STDMA and CDMA communication networks are studied. Based on asynchronous weak commitment search  the task of QoS provision is distributed among different network nodes.  Simulation results verify this scheme converges very fast to optimal solution, which makes it suitable for practical real time systems. This application of artificial intelligence in wireless and mobile communications can be used in  home automation and networking, and vehicular technology. The generalizations and extensions of this approach can be used in Long Term Evolution Self Organizing Networks (LTE-SONs). In addition, it can pave the way for decentralized and autonomous QoS provision in capillary networks that reach end nodes at Internet of Things, where central management is either unavailable or not efficient. 
\end{abstract}


\IEEEpeerreviewmaketitle

\section{Introduction}

Cognitive radio (CR) entered the lexicon of wireless communication as a means to enable dynamic spectrum access in order to increase
spectral efficiency in wireless systems. In a cognitive radio network (CRN) the spectrum license holder is called a primary user (PU) and other devices that try to dynamically access the unused resources of PU, without affecting its performance, are called secondary users or CRs. The terms secondary and CR are interchangeably used throughout the paper. 

Traffic patterns of PUs are varying, because PUs are either busy, i.e., using the link, or idle. In addition, due to inherent lower priority of CRs, they should adjust their transmission parameters to comply with PU interference requirements. In overlay CRN scheme, CRs sense the spectral resources of primary system and start transmitting when they find those to be idle, i.e., not being utilized by PU. As soon as the primary enters the band, CRs must evacuate the channel. On the other hand, in underlay scheme CRs simultaneously use the bands with primary, conditioned on avoiding interference to PUs. Throughout this article, the underlay scenario is considered.

To provide QoS in CRNs the additional constraint of avoiding interference with legacy license holders, i.e., PUs is inevitable. There are several drawbacks associated with centralized control of CRNs for QoS provision. A central management entity might not always be feasible, especially in CRNs, where the topology of the network and spectrum usage patterns are varying. Also, when the central management entity fails, the whole network experiences failure. Once a major link failure or disaster happens, the network should go to autonomous mode. Above reasons are convincing to migrate toward a distributed and autonomous management of QoS in CRNs. The distributed approach has further advantage of breaking down the load of coordination among all nodes. Here, inspired by asynchronous weak commitment search (AWCS)~\cite{yoko98}, a method for autonomous network management and recovery is put forward, which distributes the task of providing QoS among different network nodes.
 
\begin{figure}[t]
\centering
\includegraphics[width=3.5in]{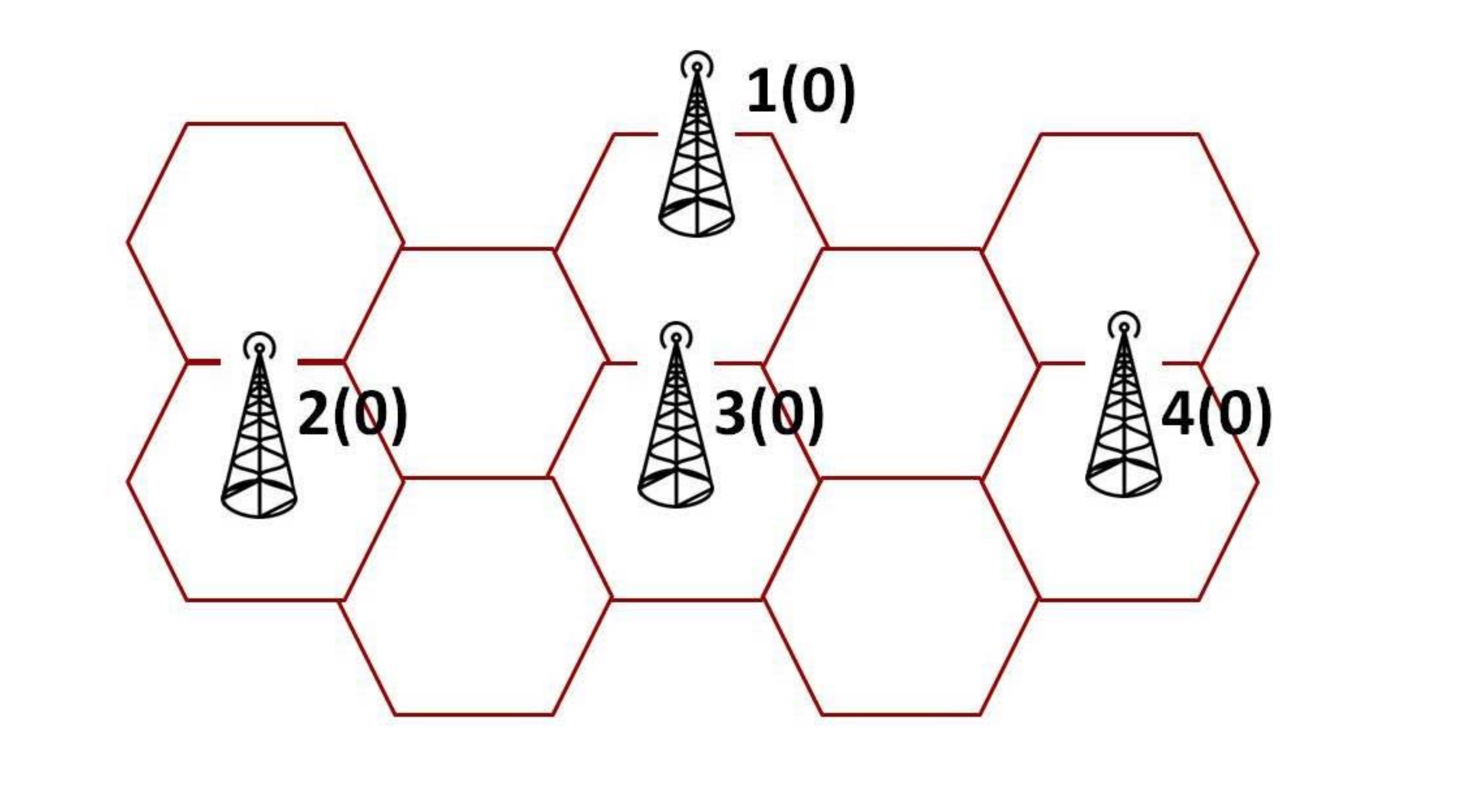}
\caption{Initial setup for example 1}
\label{fig:ex1}
\end{figure}

\begin{figure}[t]
\centering
\includegraphics[width=3.5in]{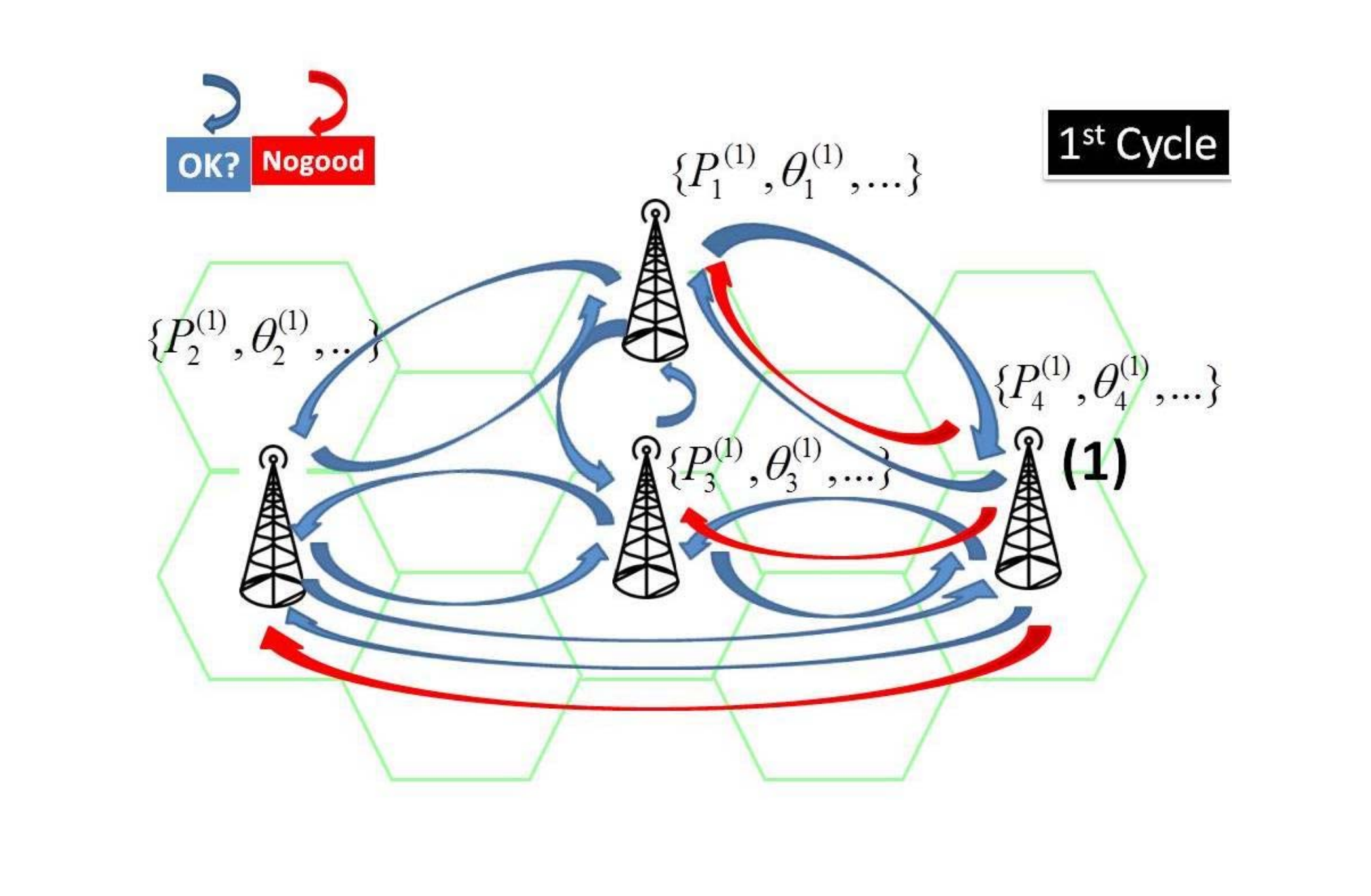}
\caption{First cycle of self-organized parameter configuration using AWCS in example 1}
\label{fig:ex2}
\end{figure}

\begin{figure}[t]
\centering
\includegraphics[width=3.5in]{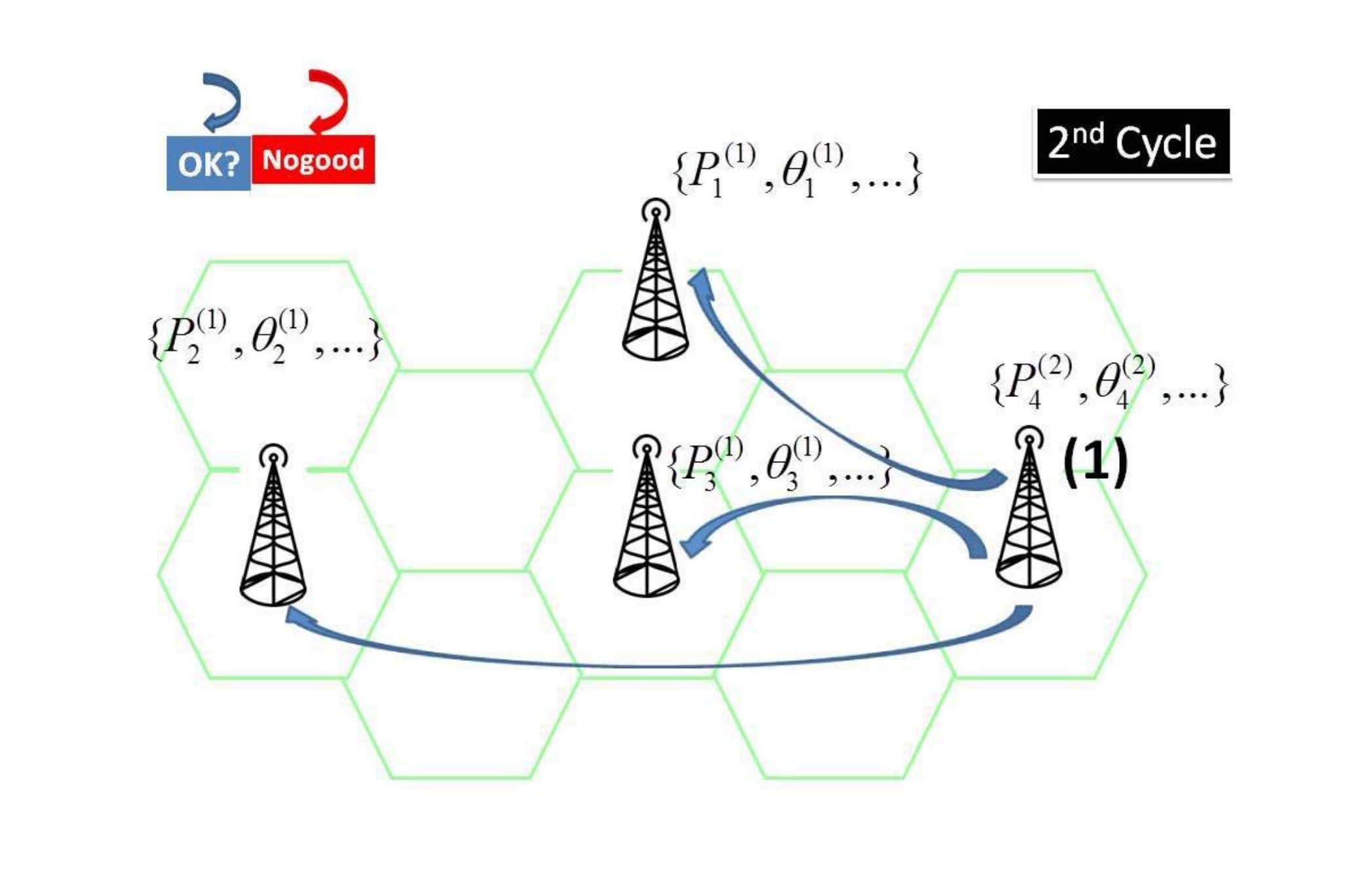}
\caption{Second cycle of self-organized parameter configuration using AWCS in example 1}
\label{fig:ex3}
\end{figure}

\begin{figure}[t]
\centering
\includegraphics[width=3.5in]{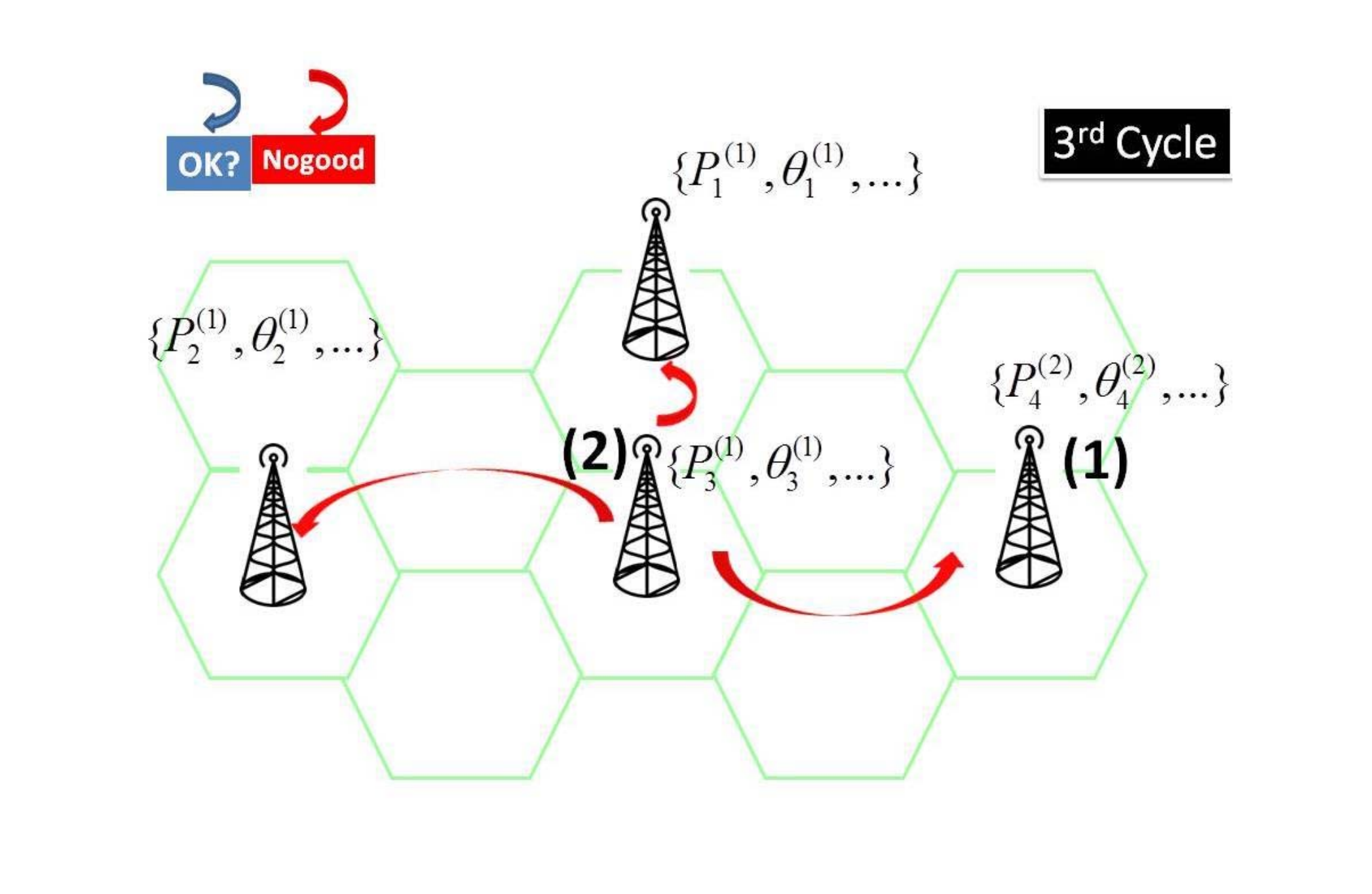}
\caption{Third cycle of self-organized parameter configuration using AWCS in example 1}
\label{fig:ex4}
\end{figure}

\begin{figure}[t]
\centering
\includegraphics[width=3.5in]{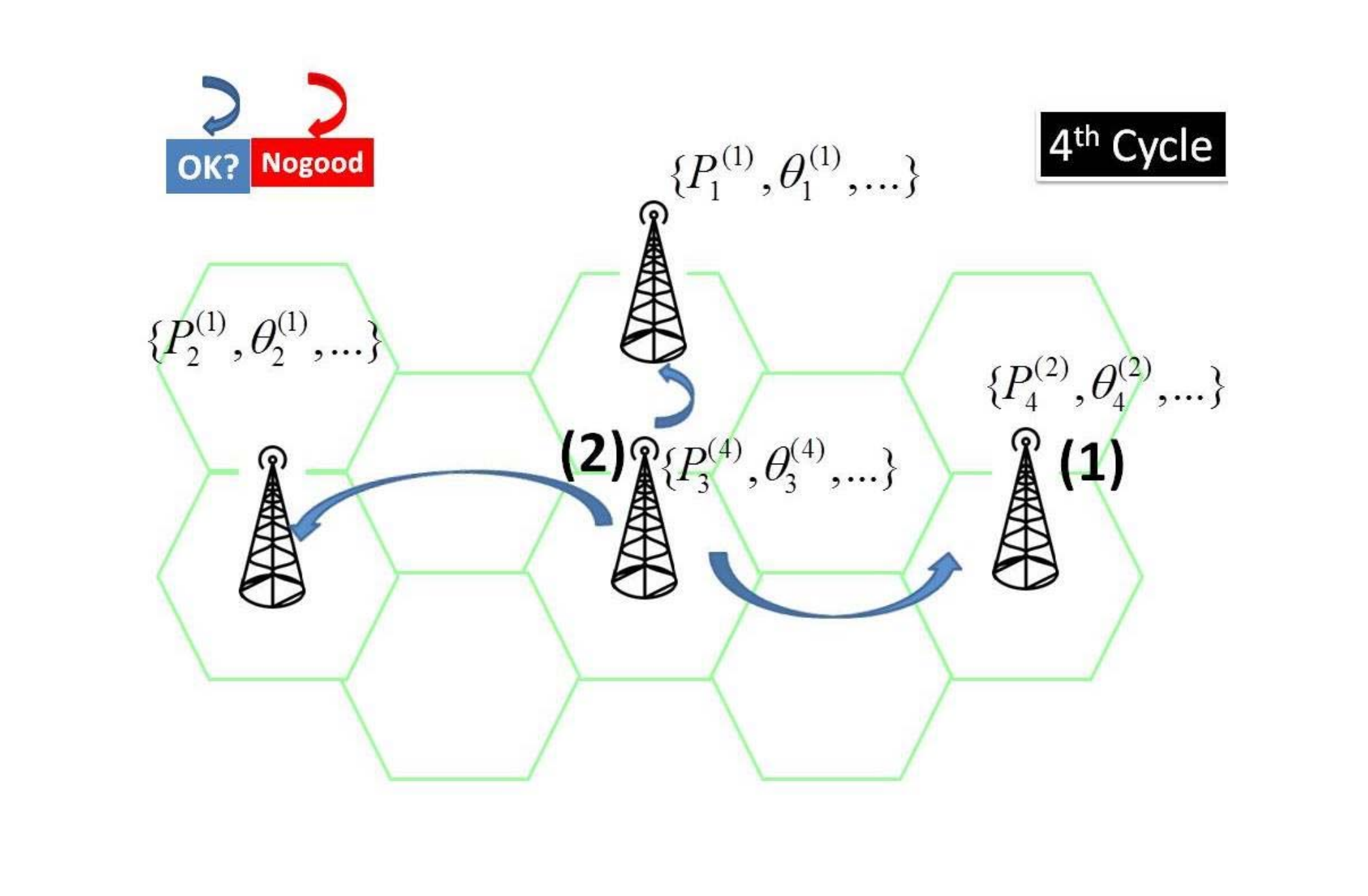}
\caption{Fourth cycle of self-organized parameter configuration using AWCS in example 1}
\label{fig:ex5}
\end{figure}

\begin{figure}[t]
\centering
\includegraphics[width=3.5in]{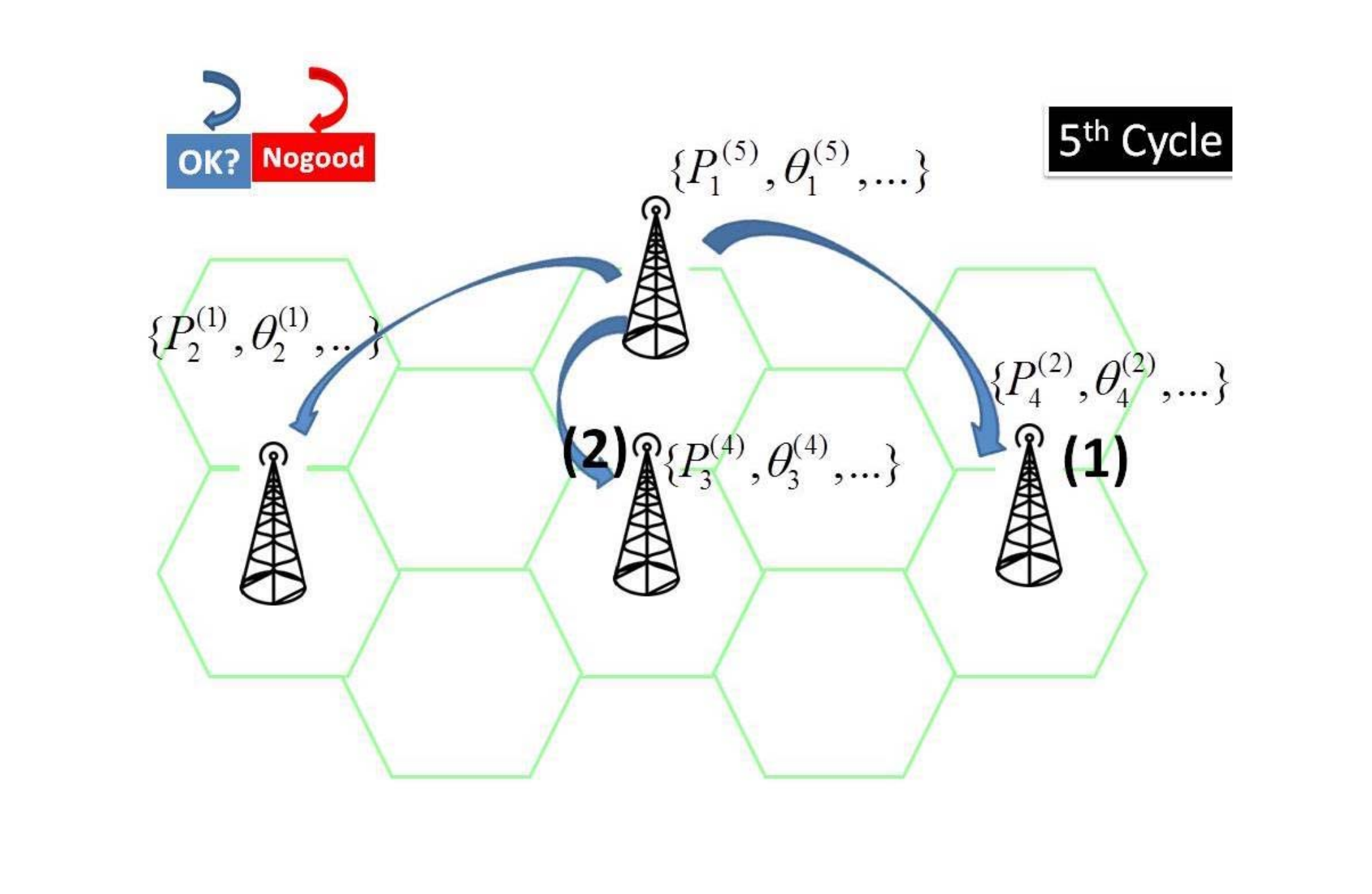}
\caption{Solution obtained at fifth cycle of self-organized parameter configuration using AWCS in example 1}
\label{fig:ex6}
\end{figure}

\textit{\textbf{Example 1}} To elucidate the many potential applications of distributed constraint satisfaction algorithms to realize self-organizing wireless communication systems, a simplified example is depicted in Figures~\ref{fig:ex1} to \ref{fig:ex6}. The method used in this example facilitates automatic self-initialization of parameters, such as antenna tilt and power in cellular networks. It can replace the tedious manual adjustments. Sets of variables and domains for each base station include power $p\in [P_{\min}, P_{\max}]$ and antenna tilt $\theta \in [\theta_{min},\theta_{\max}]$, as in Figure~\ref{fig:ex1}. Values in parentheses are priority values. Each base station selects its initial variables, and sends \textit{ok?} and \textit{nogood} messages to other base stations. Since initially all priorities are equal, the priority of CRs can be determined by a conventional order of base stations, e.g., their number. In the first cycle, shown in Figure~\ref{fig:ex2}, assume base station 4 finds constraint violations, e.g., unacceptable interference. It sends \textit{nogood} messages and increments its priority value. In the second cycle, shown in Figure~\ref{fig:ex3}, base station 4 chooses a value minimizing constraint violation, which only conflicts with base station 3. Base station 4 sends \textit{ok?} messages to other agents. In the third cycle, shown in Figure~\ref{fig:ex4}, constraint of base station 3 is violated. Therefore, base station 3 increments its priority and sends \textit{nogood} messages. 
In the 4th cycle, as in Figure~\ref{fig:ex5}, base station 3 selects a value that minimizes constraint violations, but still violating base station 1, and sends \textit{ok?} messages to other base stations. In the 5th cycle, depicted in Figure \ref{fig:ex5}, base station 1 changes its value and a solution is obtained.

The goal here is to derive a solution that satisfies all QoS constraints in the shortest possible time. To this end, AWCS is used, which is a solution to distributed constraint satisfaction problems. AWCS is an efficient method to solve such problems in comparison with other similar methods in that it offers shorter convergence time and less message exchange overhead~\cite{yokbook}.

Our protocol, as shown in Figure~\ref{fig:phas2}, is a lightweight cooperation to achieve desired QoS levels, while avoiding interference constraints of other nodes. CRs choose their values (e.g., power levels) and report it to PUs. If PUs find they are interfering, they send a one bit \textit{nogood} message, otherwise, when PUs find the SU parameter values to be in accordance with their interference constraint, they do not need to send any bit. If a SU receives a \textit{nogood} from a PU, it decreases its value and send an \textit{ok?} message again to that PU. This process is iterated until SUs make sure they are not interfering with PUs. The next step of the proposed algorithm  consists of making sure each SU does not violate the constraints of other peer SUs. In this regard, SUs use AWCS algorithm for distributed constraint satisfaction among themselves. 

This method can implement QoS in CRNs, comprising of a multitude of SUs and PUs, without the need to know mutual interference channels among nodes. In other words, the distinction of this scheme from other relevant works in the literature, e.g.,~\cite{ekr12} and \cite{ekr08}, is bypassing the need to estimate channel gains between CR transmitter receiver pairs and also between CR transmitters and PU receivers. This is due to sovling the optimization problem using distributed search.

The contributions of this article can be summarized as follows:
\begin{itemize}
\item A multi-stage protocol for decentralized QoS provision in cognitive radio networks, which utilizes AWCS at one of the stages 
\item Bypassing the need to estimate mutual interference channel gains
\item Controlled messaging overhead
\item Discussing effects of delay in passing messages on the performance
\end{itemize}

In Section~\ref{sec:body} models and constraints for QoS provision in cognitive cellular networks are presented. Section~\ref{sec:awcs} contains details of the proposed decentralized QoS protocol. Section~\ref{sec:delay} contains discussions on effects of delay of message exchange on algorithm performance. Simulation results are presented in Section~\ref{sec:simul} and Section~\ref{sec:conclus} concludes.

Table~\ref{table:not} contains the notation and abbreviations used throughout this paper.

\begin{table}
	\centering
\begin{tabular}{|l|l|}
  \hline
  CR & Cognitive radio\\
  CRN & Cognitive radio network\\
  PU & Primary user \\
  SU & Secondary user\\
  AWCS & Asynchronous weak commitment search\\
  STDMA & Spatial reuse time division multiple access\\
  CDMA & Code division multiple access\\
  SINR & Signal to interference plus noise ratio\\
  CSP & Constraint satisfaction problem\\
  $\mathcal{K}$    & Set of PU links \\
  $\mathcal{L}$   & Set of CR links\\
  $i$    & CR index \\
  $k$   & PU index \\  
  $u_i$ & Minimum Number of time slots per frame required by CR$i$ \\
  $R_i$ & Transmission rate of CR$i$\\ 
   \hline 
\end{tabular}
\caption{Notation and Abbreviation}
\label{table:not}
\end{table}

\section{System Model and Problem Statement}  
\label{sec:body} 

To show the applications of asynchronous weak commitment search algorithm in providing distributed QoS in cognitive radio networks, in the following subsections, scenarios related to cognitive underlay spatial reuse time division multiple access (STDMA) networks and cognitive underlay code division multiple access (CDMA) networks are considered. These system models involve optimal scheduling, power and rate allocations to satisfy required QoS constraints of each CR, while avoiding violation of constraints of other network nodes. 

\subsection{QoS in Cognitive STDMA Networks}

\begin{figure}[t]
\centering
\includegraphics[width=3.7in]{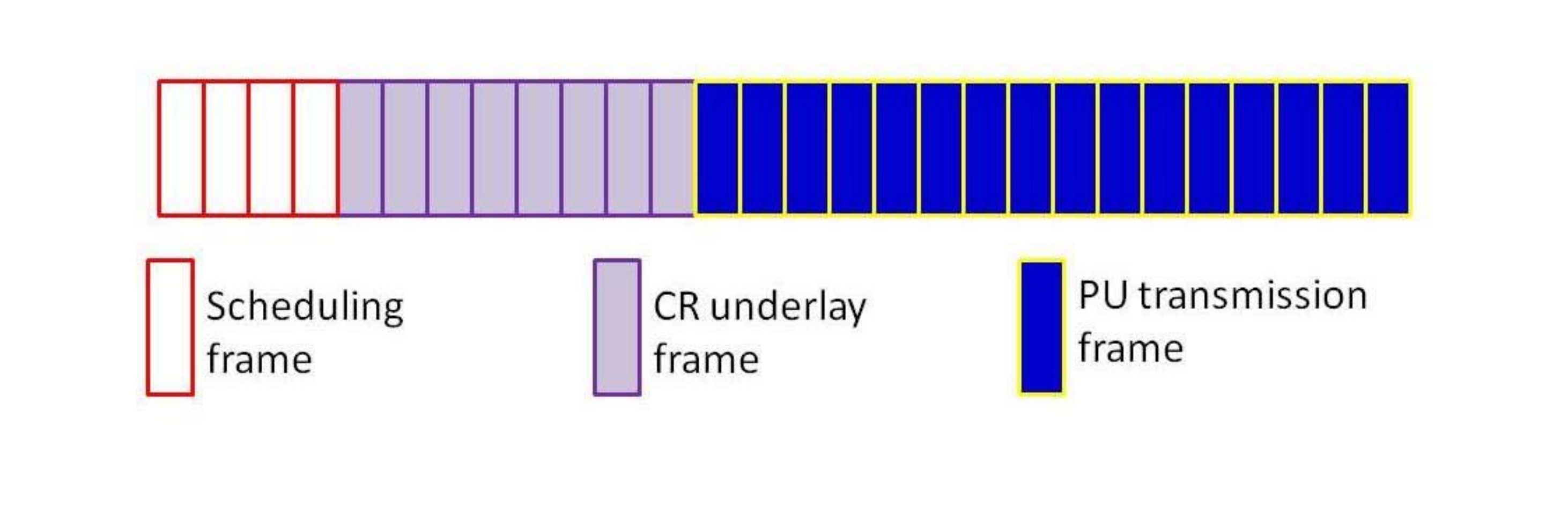}
\caption{Frame structure in STDMA cognitive transmission consisting of scheduling, underlay CR transmission, and PU transmission time slots}
\label{fig:stdm}
\end{figure}

In a cognitive underlay STDMA network a set $\mathcal{L}$ of CR links (a transmitter receiver pair), denoted by $i=1,2,\dots,|\mathcal{L}|$, coexists with a set $\mathcal{K}$ of PU links, denoted by $k=1,2,\dots,|\mathcal{K}|$~\cite{ekr12}. As shown in Figure~\ref{fig:stdm}, each frame contains a scheduling period for CR links and a transmission period. The CR transmitter receiver pairs communicate in an \textit{ad-hoc} manner. CRs use spatial resources to transmit at the same time at their scheduled time slots. One QoS requirement for each CR's traffic involves a minimum number of time-slots, $u_i$, within a frame. CRs should avoid interference to PUs, because transmission slots of CRs are underlaid with the transmissions from PUs. Using AWCS signaling messages during the scheduling period CRs find optimal schedules and transmission powers to be used during the transmission period.

The other required QoS constraint is the signal to interference plus noise (SINR) level at the receiver, which must be above a certain threshold for the data to be transmitted successfully over a link. To this end, path loss should be taken into consideration. Without loss of generality, several propagation models can be assumed for path loss. Examples include indoor, outdoor, and deterministic propagation models. 

\begin{figure}
\begin{flushleft}\includegraphics[width=4.3in]{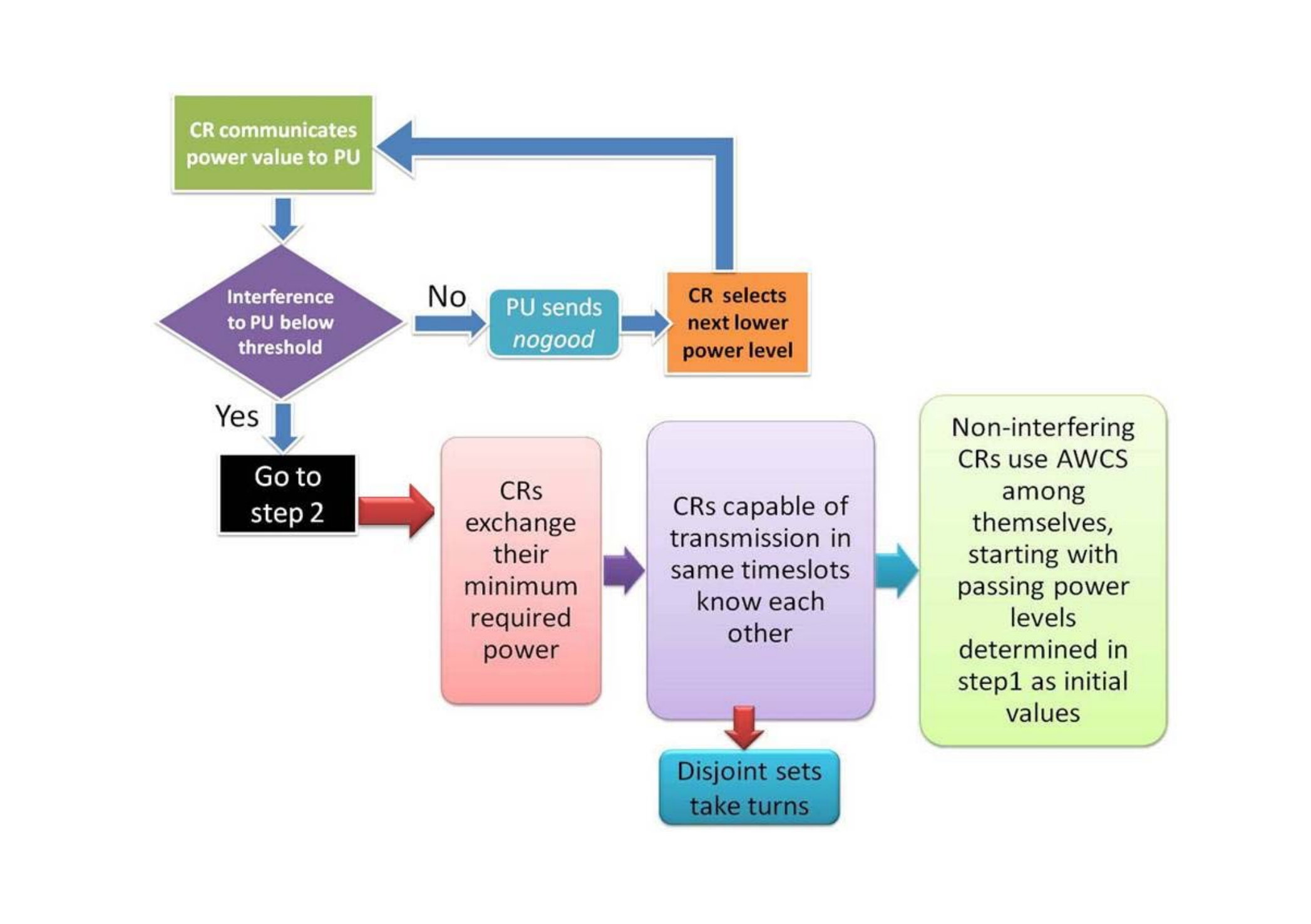}
\caption{Steps of our suggested decentralized QoS protocol for cognitive STDMA networks}
\label{fig:phas2}
\end{flushleft}
\end{figure}

Within a time-slot a subset of the CR links $\mathcal{L}$ can simultaneously transmit data, i.e., have transmission powers greater than zero, to spatially reuse the frequencies, provided that they meet the interference limit constraints of PUs and the minimum SINR requirements of CR links~\cite{ekr12}.  $\mathcal{S}$ denotes the set of all feasible access patterns of CR links indexed by $s=1,2,\dots,|\mathcal{S}|$. Binary matrix $\mathbf{Q}$ of size $\vert \mathcal{L} \vert \times \vert \mathcal{S}\vert$ contains the activity of all CR links in all feasible access patterns $\mathcal{S}$. 
\begin{align}
q_{i,s} =&\begin{cases} 1 & \mbox{if } n\mbox{ CR}i \mbox{ is active in~}  \mathcal{S}_s \\ \nonumber 0 & \mbox{otherwise. }  \end{cases}\\ 
&\forall q_{i,s}\in \mathbf{Q}
\end{align} 
 
CRs should meet their traffic demands $u_i$, and at the same time minimize their transmission length in terms of number of time slots  $\vary_s$ dedicated to a particular access pattern within a frame. As mentioned before, their SINR constraints and SINR constraints of PUs also hold. On the other hand, the power budget of CRs is limited~\cite{ekr12}. 

\subsection{Qos for Cognitive CDMA Cellular Networks}

\begin{figure}[b]
\centering
\includegraphics[width=3.7in]{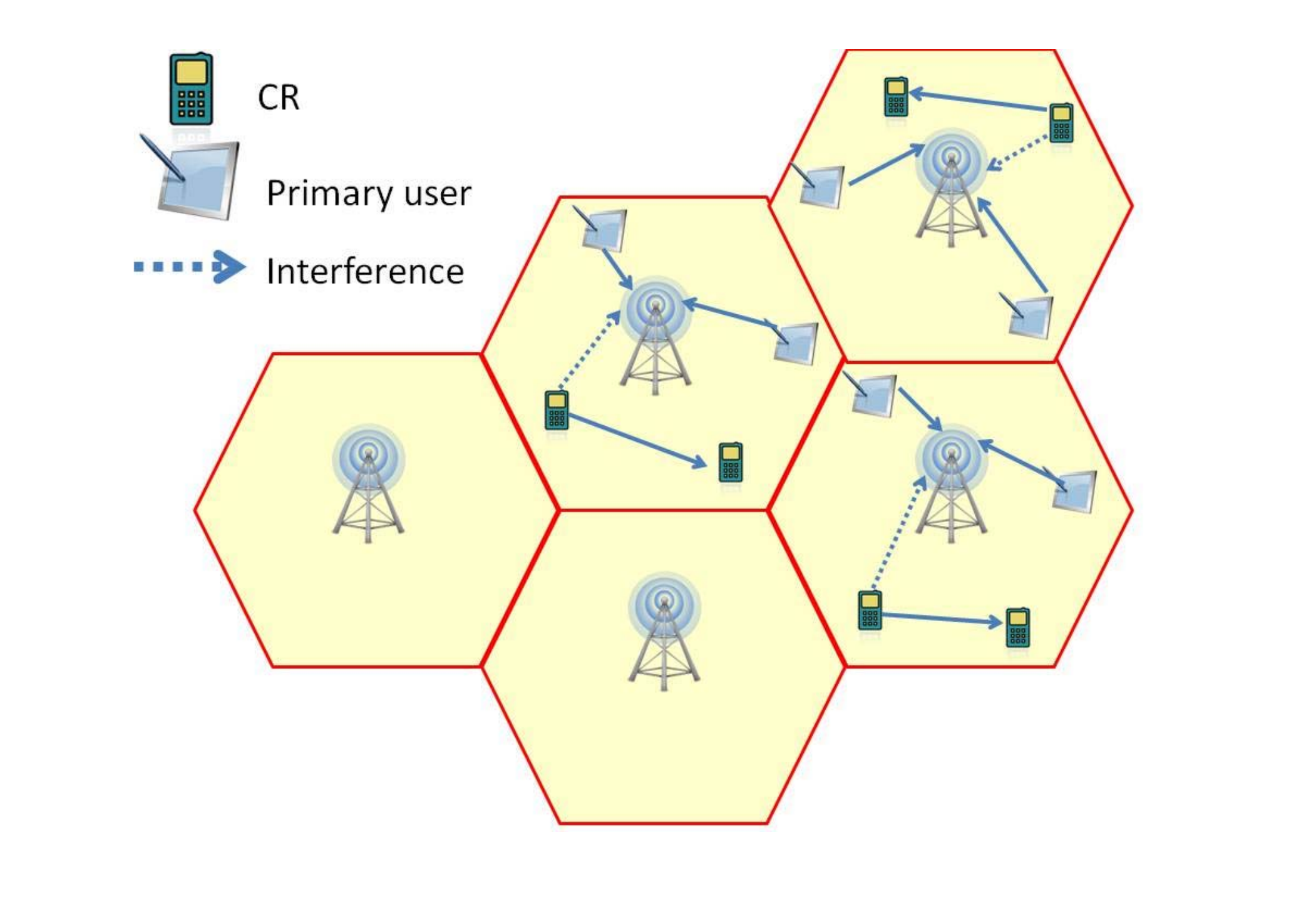}
\caption{CDMA cellular network with coexisting primary users and CRs}
\label{fig:cdma}
\end{figure}

Here, providing QoS in terms of rate and power allocation for cognitive CDMA cellular networks~\cite{ekr08} is considered, where PUs and CRs can transmit simultaneously in a shared frequency band. The scenario is shown in Figure~\ref{fig:cdma}, where PUs in a cellular wireless network communicate with the corresponding base stations~\cite{19ekr8}. Rates and power values selected by CRs should satisfy QoS requirements in terms of SINR and minimum data rates, while at the same time the interference to primary base stations must be kept below a certain threshold.

In contrast with~\cite{ekr08}, where it is assumed that a central controller in the CR network performs the joint admission control, and rate/power allocation for CRs, a decentralized method using distributed constraint satisfaction is deployed. As a result, our method does not need knowledge of channel gains among PUs and CRs and peer CRs.

The goal is to maximize the proportionally fair data rates $\sum_{i=1}^{|\mathcal{L}|} \ln (R_i)$ of CRs~\cite{ekr21}. The problem involves finding fair resource allocations for CRs subject to interference constraints introduced to PUs in a cognitive cellular CDMA network~\cite{ekr08}.

The transmission rate of each CR should fall within the allowed maximum and minimum rate values $R_{\max}$ and $R_{\min}$, respectively. The power budget available to CRs is limited and interference to PUs, resulting from CRs underlay transmission, should be kept below the desired level. Also, each CR has a minimum SINR threshold for its transmitted data to be successfully received.

To keep interference to PUs below the desired threshold, the PU informs CRs that their selected transmit powers are above threshold. Accordingly, CRs decrease their power to the next lower quantized level until the interference constraint of PU is met. Nevertheless, if through some mechanisms, channel gains or their statistics can be estimated by CR nodes, the algorithm would converge faster to a solution, because of the insight into selection of initial values. One possible scheme for estimation of average channel gains can be using pilot signals and channel reciprocity owing to using same frequencies. 

\section{QoS Using Asynchronous Weak Commitment Search for Distributed Constraint Satisfaction}
\label{sec:awcs}

A constraint satisfaction problem (CSP) consists of $n$ variables $x_1, x_2, ..., x_n$, whose values are taken from finite, discrete domains $D_1, D_2,\dots,D_n$, respectively, and a set of constraints on their values~\cite{yoko98}. Solving a CSP is equivalent to finding an assignment of values to all variables such that all constraints are satisfied.

In a distributed CSP variables and constraints are distributed among autonomous agents. Agents that are related by constraints are called neighbors and communicate by sending messages. The random delay in delivering a message is finite. Furthermore, for the transmission between any two agents, messages are received in the order in which they were sent.

Each agent records its own \textit{agent\textendash view} and \textit{nogood}~\cite{vidal-sync}. Here, agents are associated to different CR nodes. The \textit{agent\textendash view} of a CR is the set of values (e.g., transmission power) selected by other CRs and communicated to it. A \textit{nogood} is a subset of \textit{agent\textendash view}. If a \textit{nogood} exists, it means the agent cannot find a value from
the domain of its variable to be consistent with the \textit{nogood}. When the \textit{agent\textendash view} of a node contains a \textit{nogood}, the values of other agents must be changed. CRs exchange assignments and \textit{nogoods}. When a CR receives new variable assignments initiated by other CRs, it updates its \textit{agent\textendash view} accordingly. A \textit{nogood} received by a CR is accepted if it is consistent with its \textit{agent\textendash view}, otherwise it is discarded. When a CR cannot take any value consistent with its \textit{agent\textendash view}, because of the original constraints or because of received \textit{nogoods}, new \textit{nogoods} are generated as inconsistent subsets of the \textit{agent\textendash view}, and sent to other CRs. The process terminates when a solution has been found, or when the empty \textit{nogood} is generated, which means the problem has no solution~\cite{vidal-sync}.

AWCS is an efficient algorithm for solving distributed constraint satisfaction problems involving multiple agents. AWCS algorithm uses the two types of \textit{ok?} and \textit{nogood} messages, with the same significance.
When an agent receives an \textit{ok?} message, it updates its \textit{agent\textendash view} list and checks if its constraints are violated. If no \textit{nogood} value of higher priority agents is violated, no action is required. If there are a few higher priority \textit{nogood} values that have inconsistent values and these values could be eliminated by changing the variable assignment, the CR will change this value and will send the \textit{ok?} message~\cite{vidal-sync},~\cite{yoko98}. 
\textit{Nogood} learning can improve performance of AWCS algorithm, in terms of required cycles to solve the problem~\cite{learn}.

AWCS algorithm uses the min-conflict heuristic as a value ordering heuristic.  In min-conflict heuristic when selecting a variable value, if there exist multiple values consistent with the \textit{agent\textendash view}, i.e., those that satisfy the constraints with variables of higher priority CRs, the agent prefers the value that minimizes the number of constraint violations with variables of lower priority agents. Using this method without any CR having exact information on the partial solution, CRs can operate concurrently and asynchronously. 

For each CR a non-negative integer value representing the priority order of the CR is defined. This is called the \textit{priority value}. Any CR with a larger \textit{priority value} has higher priority. In case the \textit{priority values} of multiple CRs are the same, the order is determined by an agreed upon convention. For each CR the initial \textit{priority value} is 0. If there exists no consistent value for CR$i$, the priority value of CR$i$ is changed to $l+1$, where $l$ is the largest \textit{priority value} among CRs.

  Assuming CRs agree, before the protocol start, on their priority orders, using AWCS, CRs can revise a violating value assignment decision without an exhaustive search by dynamically changing the priority order of CRs\cite{yokbook}. In other words, when a CR cannot find a value consistent with the higher priority CRs, the priority order is changed to yield that CR the highest priority. Therefore, when a CR makes a mistake in selecting a value for its variables, e.g., transmit power level, the priority of another CR becomes higher and consequently, the CR that made the bad decision will not commit to it, and the selected value is changed. This implies giving up the partial solution if there exists no consistent value with the partial solution and restarting the search process. This is obtained through dynamically changing the priority order. 

Assuming equal rates for CRs, the details of using AWCS for QoS provision in cognitive CDMA networks is described in Algorithm 1. 
  
\begin{algorithm}
  \caption{Distributed power allocation by CR$i$ for QoS provision in a cognitive CDMA network with equal rates using single variable AWCS~\cite{yokbook}}
  \begin{algorithmic} \footnotesize
    \State \textbf{when received} (\textbf{ok?}, (CR$j$, $p_j$, \textit{priority})) \textbf{do}
    \State  add (CR$j$, $p_j$, \textit{priority}) to \textit{agent\textendash view};
    \State \textbf{check\textendash agent\textendash view}; \textbf{end do}\\
    \State \textbf{when received} (\textbf{nogood}, CR$j$, \textit{nogood}) \textbf{do}
    \State add \textit{nogood} to \textit{nogood\textendash list};
    \State \textbf{when} (CR$k$, $p_k$, \textit{priority}) where CR$k$ is not in \textit{neighbors} is contained in \textit{nogood} \textbf{do}
    \State add CR$k$ to \textit{neighbors}, add (CR$k$, $p_k$, \textit{priority}) to \textit{agent\textendash view}; \textbf{end do};
    \State \textbf{check\textendash agent\textendash view};\textbf{end do};\\
   \State procedure \textbf{check\textendash agent\textendash view}
   \State \textbf{when} \textit{agent\textendash view} and \textit{current\textendash value} are not consistent \textbf{do}
   \State \textbf{if} no value in $D_i$ is consistent with \textit{agent\textendash view} \textbf{then backtrack};
    \State \textbf{else} select $p\in D_i$ where \textit{agent\textendash view} and $p$ are consistent and $p$ minimizes the number of constraint violations with lower priority CRs
\State \textit{current\textendash value} \textleftarrow $p$;
  \State send (\textbf{ok?},(CR$i$, $p$, \textit{current\textendash priority})) to \textit{neighbors}; \textbf{end if}; \textbf{end do};\\
\State procedure \textbf{backtrack}
\State \textit{nogoods} \textleftarrow $\{V|V=\mathrm{inconsistent~subset~of~}\mathit{agent\textendash view}\}$;
\State \textbf{when} an empty set is an element of \textit{nogoods} \textbf{do}
\State broadcast to other CRs that there is no solution,
\State terminate this algorithm; \textbf{end do};
\State \textbf{when} no element of \textit{nogoods} is included in \textit{nogood\textendash sent} \textbf{do} 
\State for each $V\in$ \textit{nogoods} \textbf{do}
\State add $V$ to \textit{nogood\textendash sent}
\State \textbf{for each} (CR$j$, $p_j$, $priority(j)$) in $V$ \textbf{do};
\State send (\textbf{nogood}, CR$i$, $V$) to CR$j$; \textbf{end do}; \textbf{end do};
\State $priority(\max)$ \textleftarrow $\max_{(CRj,~p_j,~priority)\in \mathit{agent\textendash view}} (priority(j)$)
\State \textit{current\textendash priority}\textleftarrow $1+priority(\max)$
\State select $p\in D_i$, where $p$ minimizes the number of constraint violations with lower priority CRs;
\State \textit{current\textendash value} \textleftarrow $p$
\State send (\textbf{ok?},(CR$i$, $p$, \textit{current\textendash priority})) to \textit{neighbors}; \textbf{end do};
  \end{algorithmic}
\end{algorithm}

\subsection{Decentralized QoS Provision in Cognitive CDMA Networks with Unequal Rates}

In the above, it was shown how AWCS can be used for optimal power allocation among CRs in CDMA networks with equal rates. Here, a modified version of AWCS is used related to when CRs have multiple local variables to optimize~\cite{yokbook}, e.g., rate and power. Algorithm 2 contains procedures executed by CR$i$ with two variables of rate and power when receiving \textit{ok?} messages. It is an AWCS algorithm for multiple local variables~\cite{yokbook} and originates from AWCS for one local variable, but a CR sequentially performs the computation for each variable, and communicates with other CRs only when it can find a local solution that satisfies all local constraints. By using this algorithm, a bad local solution can be modified without forcing other CRs to exhaustively search their local solutions, and the number of interactions among CRs can be decreased. This algorithm is more efficient than an algorithm that employs the prioritization among agents and a simple extension of AWCS for the case of a single local variable~\cite{yokbook}.
 
\begin{algorithm}\footnotesize
  \caption{\textit{ok?} messages in distributed rate and power allocation for QoS provision in cognitive CDMA networks with unequal rates using multi-variable AWCS~\cite{yokbook}}
  \begin{algorithmic}
    \State \textbf{when received} (\textbf{ok?}, (\textit{sender\textendash id, variable\textendash id, variable\textendash value, priority})) \textbf{do}
    \State ~add (\textit{sender\textendash id, variable\textendash id, variable\textendash value, priority}) to \textit{agent\textendash view};
    \State \textbf{when} \textit{agent\textendash view} and \textit{current\textendash assignments} are not consistent \textbf{do}
    \State ~~\textbf{check\textendash agent\textendash view}; \textbf{end do};\\
    \State procedure \textbf{check\textendash agent\textendash view}
    \State \textbf{if} \textit{agent\textendash view} and \textit{current\textendash assignements} are consistent \textbf{then} 
    \State communicate changes to related CRs;
    \State \textbf{else} select $x_k$, which has the highest priority and violating some constraint with higher priority variables;
   \State \textbf{If} no value in $D_k$ is consistent with \textit{agent\textendash view} and \textit{current\textendash assignments} 
   \State \textbf{then} record and communicate a nogood, i.e., the subset of \textit{agent\textendash view} and \textit{current\textendash assignments}, where $x_k$ has no consistent value;
   \State \textbf{when} the obtained nogood is new \textbf{do} 
    \State set $x_k$'s priority value to the highest priority value of related variables + 1;
\State select $d \in D_k$ where $d$ minimizes the number of constraint violations with lower priority variables;
  \State set the value of $x_k$ to $d$;
\State \textbf{check\textendash agent\textendash view}; \textbf{end do};
\State \textbf{else} select $d\in D_k$ where $d$ is consistent with \textit{agent\textendash view} and \textit{current\textendash assignments}, and minimizes the number of constraint violations with lower priority variables;
\State set the value of $x_k$ to $d$;\\
\textbf{check\textendash agent\textendash view}; 
\State \textbf{end if};
\State \textbf{end if};
  \end{algorithmic}
\end{algorithm}

Every CR changes the values of its local variables in order. It selects a variable that has the highest priority among variables that are violating constraints with higher priority variables, and modifies its value so that constraints with higher priority variables are satisfied. When all local variables satisfy constraints with higher priority variables, the CR sends changes to other CRs.

After CRs choose initial values by starting from the maximum allowed power levels, each CR communicates these initial values via \textit{ok?} messages.  After that, CRs wait for and respond to messages they receive. Any CR can handle multiple messages concurrently, i.e., it first revises \textit{agent\textendash view} according to messages, and performs $\textbf{check\textendash agent\textendash view}$ only once. 
By sending messages to other CRs only when a CR finds a consistent local solution, the number of messages exchanged among CRs decreases. 

In our proposed protocol, the interaction among CRs is different from their interaction with PUs. This is due to the fact that, in CRN paradigm, PUs should neither be affected by interference nor adjust their values according to CRs, but the other way around. Therefore, our protocol has two phases. In the first phase, CRs send their values to PUs. PUs, on the other hand, check if the value selection of CRs violate their QoS requirements by exceeding the interference threshold. If not, they do not send any messages. However, if they find the value selection of CRs to be violating, they send a one bit message to the corresponding CR. The CR then decreases its selected value to next quantized lower level and iterates the process until the constraints of PUs are met. CRs then enter the second phase to satisfy the constraints among themselves by using AWCS algorithm. 

\subsection{QoS Solution for Cognitive STDMA Networks}

As shown in Figure~\ref{fig:phas2}, CRs first exchange their minimum required SINR levels to be able to find out, right from the beginning, if the problem has a solution. If the required SINR level of CRs does not result in mutual interference, no messages are exchanged. Otherwise, a one bit message is sent from the victim CR. This way, CRs find out which groups of them can use same time slots for transmission. Then, each group of non-interfering CRs uses AWCS to find out what maximum possible power levels they can use, besides each other, while keeping it below the power budget of each CR and still causing no interference to other CRs. They start with selecting their maximum power budget and if a constraint violation occurs, they decrement their selected power to the next lower level. This process iterates by exchanging \textit{ok?} and \textit{nogood} messages until AWCS terminates with a solution. After adjusting power levels, interfering subsets of CRs should schedule transmission time slots based on their traffic demands and interference. For this part of our protocol it is suggested that CRs use a conventional order (e.g., using a number assigned to each CR) at the beginning of the first frame. Disjoint sets of interfering CRs, not allowed to use same time slots within a frame, form a partition. Each set is specified by its head, i.e., the CR in that set with lowest identification number. At each frame sets update their order in a circular fashion. Within a single frame, the set with lowest head number uses its required time slots, then, the next set and so forth. For next frame, in a circular round robin fashion, the set whose head had the largest number among all heads, now goes in the first place and the set with lowest head number, which was first in previous frame takes the second position. Within this frame, CRs in the first set have priority to use time slots according to their minimum traffic requirements and the second priority belongs to the set in the second place to start transmitting in required time slots. 

To shed light on this step of the suggested protocol, consider, for instance, 7 CRs partitioned into 3 disjoint sets, based on interference, i.e., $\{CR1,CR2,CR4\}, \{CR3,CR6\}, \{CR5,CR7\}$. In other words, $CR5$ and $CR7$ are spatially apart so they can use same frequencies for transmission in same time slots. The heads are $CR1$, $CR3$ and $CR5$. In the first frame, the set containing $CR1$ is scheduled first in using its required time slots, based on its users' minimum traffic demands. Then, the set containing $CR3$, and at last, $CR5$ and $CR7$ get to use time slots. In the second frame, the priority of sets, to use time slots, changes in a circular round robin fashion. It means, the new order of heads is $CR5$, $CR1$, $CR3$. In other words, $CR5$ and $CR7$ have priority over other CRs. After they use their time slots, the set of $\{CR1,CR2,CR4\}$ is scheduled and finally, the set containing $CR3$ and $CR6$. In the third frame, the priority will belong to the set whose head is $CR3$.    

Next, the impact of message delays on the decentralized QoS provision protocol is investigated.

\section{Effects of Delay in Message Exchange}
\label{sec:delay}

In practice, messages do not arrive instantaneously but are delayed due to network properties. Asynchronous distributed constraint satisfaction techniques have different behaviors when delays happen in exchanging messages~\cite{vidal-sync}. Delays can vary on account of different network factors, such as hardware and topology. 
Effects of message delays on distributed constraint satisfaction algorithms have been measured using controlled simulation environments that apply randomly generated delays~\cite{amnon}.  The main results, with respect to application in CRNs are briefly described.

In simulation based delay analysis, agents (or threads) run asynchronously, exchanging messages by using a common mailer~\cite{amnon}. The mailer can simulate message delays, but, needs to be controlled by an algorithm that takes into account the concurrent time-keeping of the asynchronous system. 

The mailer holds a global Logical Time Counter (LTC) of concurrent computation steps performed by agents in the system. Every message delivered by the mailer to an agent carries the LTC value of its delivery to the receiving agent.
When the mailer receives a message, it first checks the LTC value that the message carries. If it is larger than its own value, it increments the value of the LTC. This updates the global clock of the Mailer, which is the largest of all the logical times of all agents. By incrementing  LTC only when messages carry LTCs with values larger than the mailer's LTC value, steps that were performed concurrently are not counted twice~\cite{amnon}.

For asynchronous algorithms, such as AWCS, two messages between the same pair of agents must arrive in the same order they were sent.

Based on the number of concurrent assignments by agents, distributed search algorithms for constraint satisfaction are divided into two categories, i.e., single search process algorithms and concurrent (multiple) search process algorithms. AWCS is a single process algorithm in that all the variables in it have exactly one assignment at each instant of its run~\cite{amnon}. In concurrent or multiple search algorithms for solving distributed constraint satisfaction problems the search space is broken down into disjoint subspaces on which concurrent processes perform their search~\cite{55amnon}.

There exist two criteria to evaluate the algorithms. One is the total number of nonconcurrent constraint checks (NCCCs) and the other is the total number of exchanged messages.

AWCS reads multiple messages at each step, based on their instantaneous arrival. 

Since in concurrent search algorithms agents perform computation against consistent partial assignments, performance of such algorithms is less affected by message delay in terms of the load on the network, and number of messages. In other words, computation performed in one search sub-space, while others are delayed is not wasted as in a single search process, such as AWCS.

In AWCS agents perform assignments asynchronously and when the updating message is randomly delayed, some of their computation can be irrelevant due to inconsistent \textit{agent\textendash views}~\cite{54amnon}. This is a consequence of the fact that with random message delays, agents might respond to a single
message, instead of all messages sent in the previous cycle, and since messages in AWCS are sometimes conflicting, agents perform more unnecessary computation steps when responding
to fewer messages in each cycle. Hence, the improvement that results from reading all incoming messages in each step~\cite{61amnon}, which is intrinsic to AWCS, is no longer useful in case of random message delays. Therefore, the number of NCCCs and messages sent in AWCS grow with the size of the message delay.

From above, it is concluded that there exits a tradeoff between selecting either a non-concurrent algorithm, such as AWCS, or a concurrent algorithm. When delays happen, concurrent algorithms perform better, however, concurrent algorithms require each CR to be equipped with more processing power to handle the concurrent computations, in terms of breaking the variable search space into multiple sub-spaces and keeping track of multiple assignments to a variable all at the same time. 

\section{Performance Evaluation and Analysis}
\label{sec:simul}

Simulations include a primary network coexisting with an \textit{ad-hoc}
CRN. In deterministic path loss model channel gains can be expressed as the product of a spreading gain $B$ and a negative exponent of the distance $d$, e.g., $B^{-1} d^{-4}$~\cite{util}. 

Performance is analyzed in terms of the number of cycles required to solve the problem, number of messages exchanged, and allocated resource levels, considering various determining parameters, such as power, quantization levels, SINR threshold, number of nodes, \textit{etc}.

\begin{figure}[t]
\centering
\includegraphics[width=3.7in]{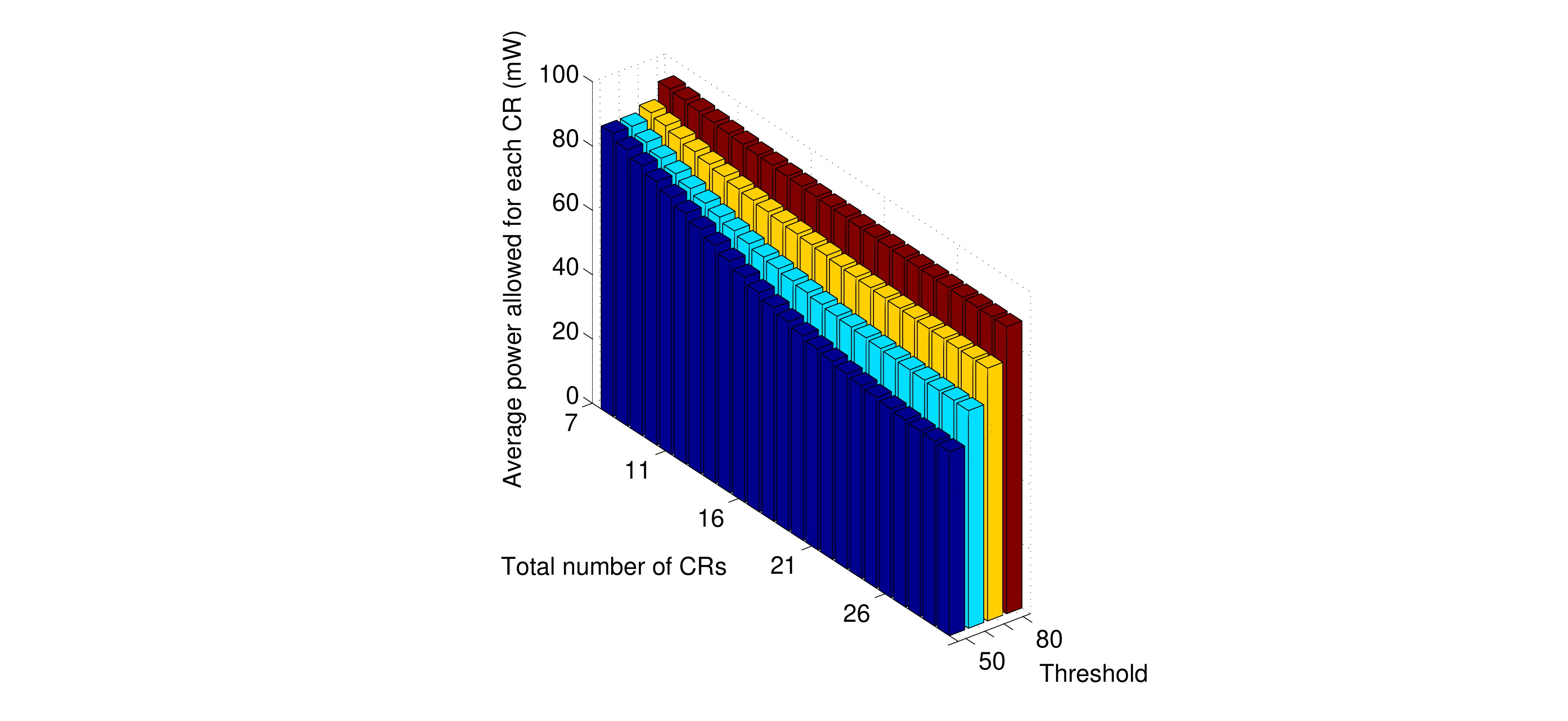}
\caption{Average allocated power for each CR vs. total number of CRs and interference threshold}
\label{fig:pow1}
\end{figure}

\begin{figure}[t]
\centering
\includegraphics[width=3.7in]{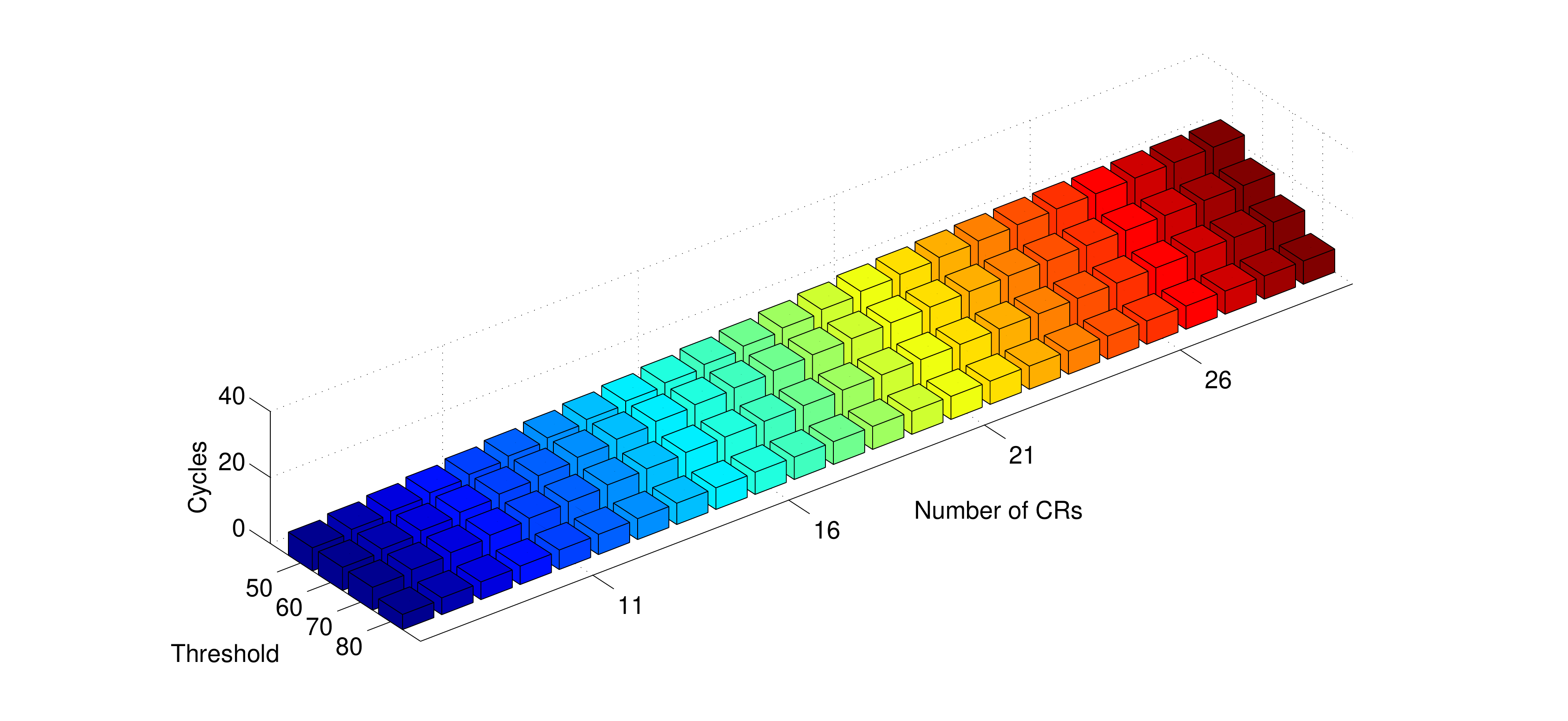}
\caption{Cycles vs. total number of CRs and interference threshold}
\label{fig:cycle1}
\end{figure}

\begin{figure}[t]
\centering
\includegraphics[width=3.7in]{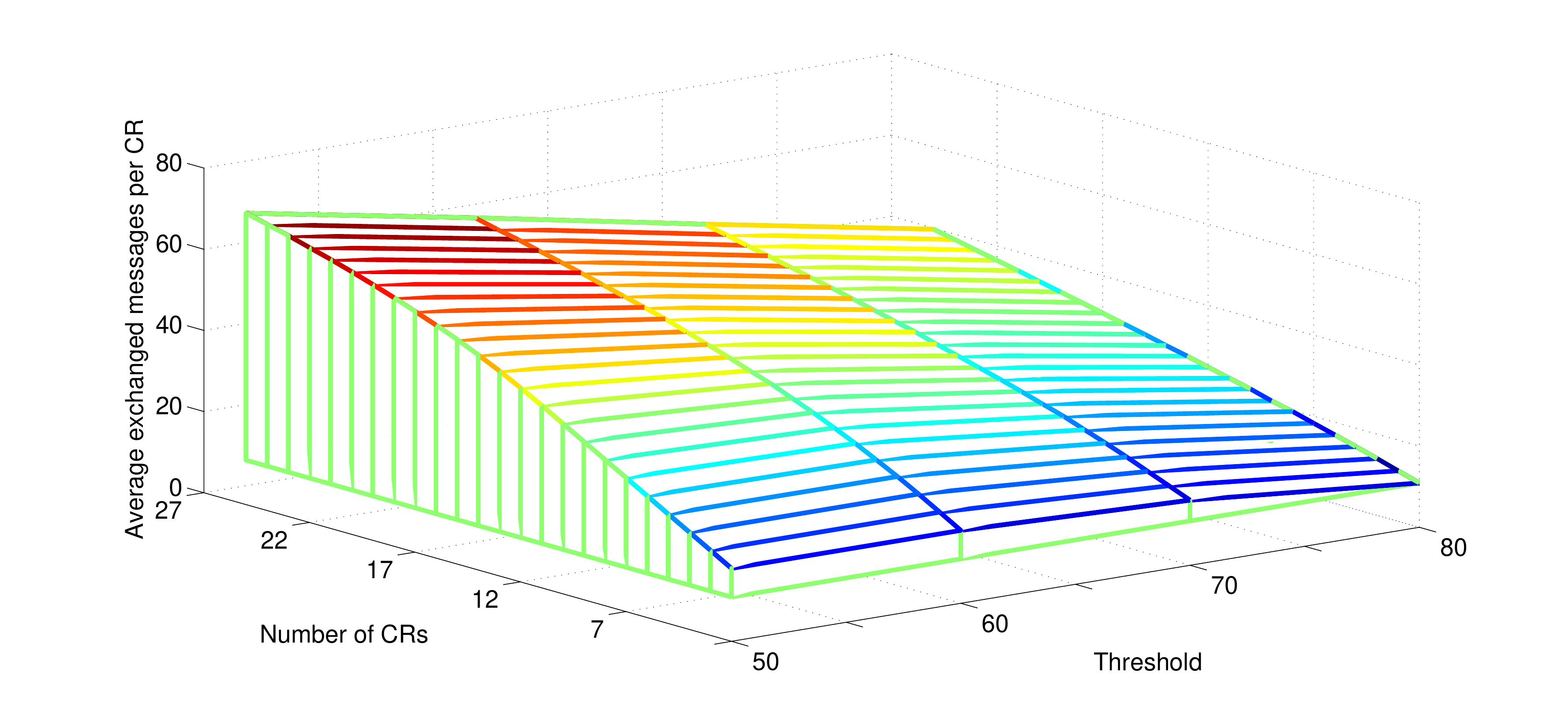}
\caption{Exchanged messages per CR vs. threshold and total number of nodes}
\label{fig:msg1}
\end{figure}

\begin{figure}
\centering
\includegraphics[width=3.7in]{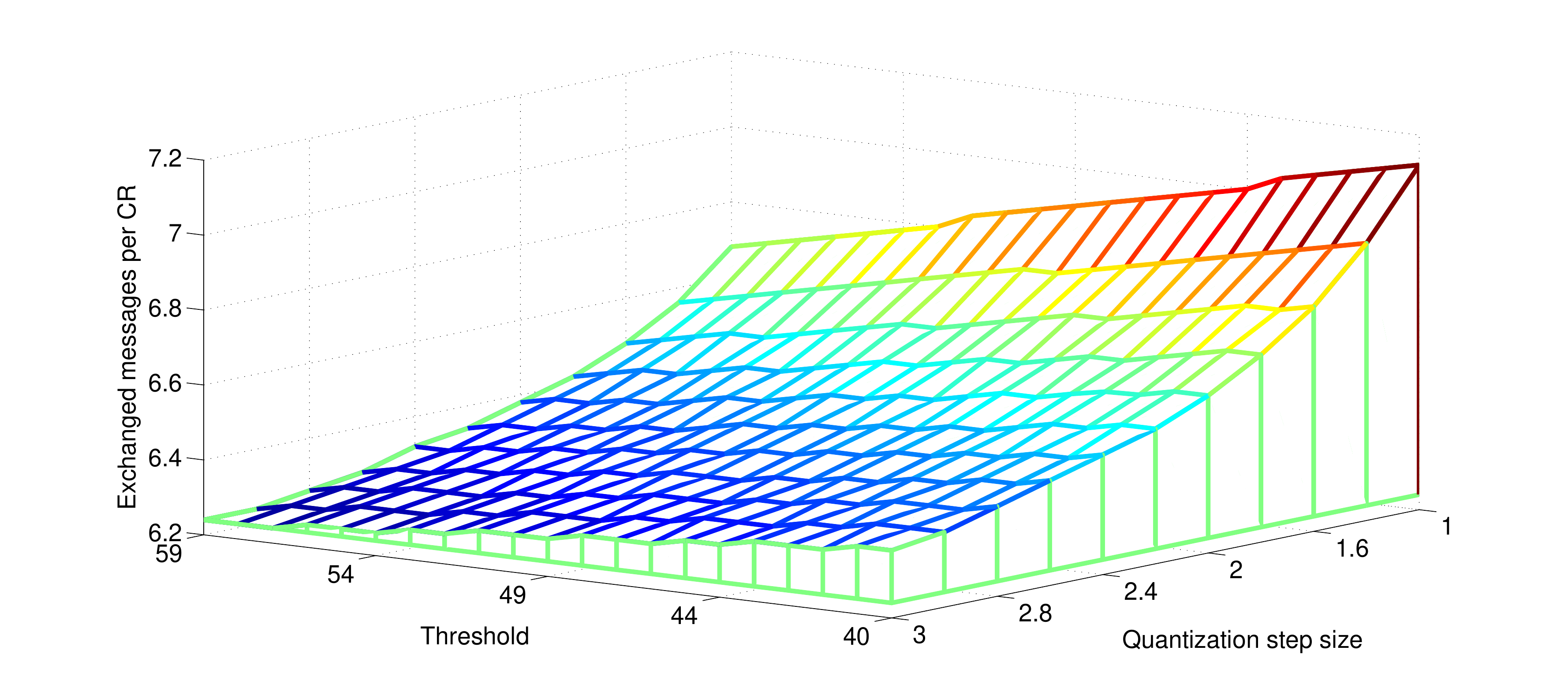}
\caption{Exchanged messages vs. power quantization step size and threshold for 7 CRs}
\label{fig:quant}
\end{figure}

\begin{figure}[t]
\centering
\includegraphics[width=3.7in]{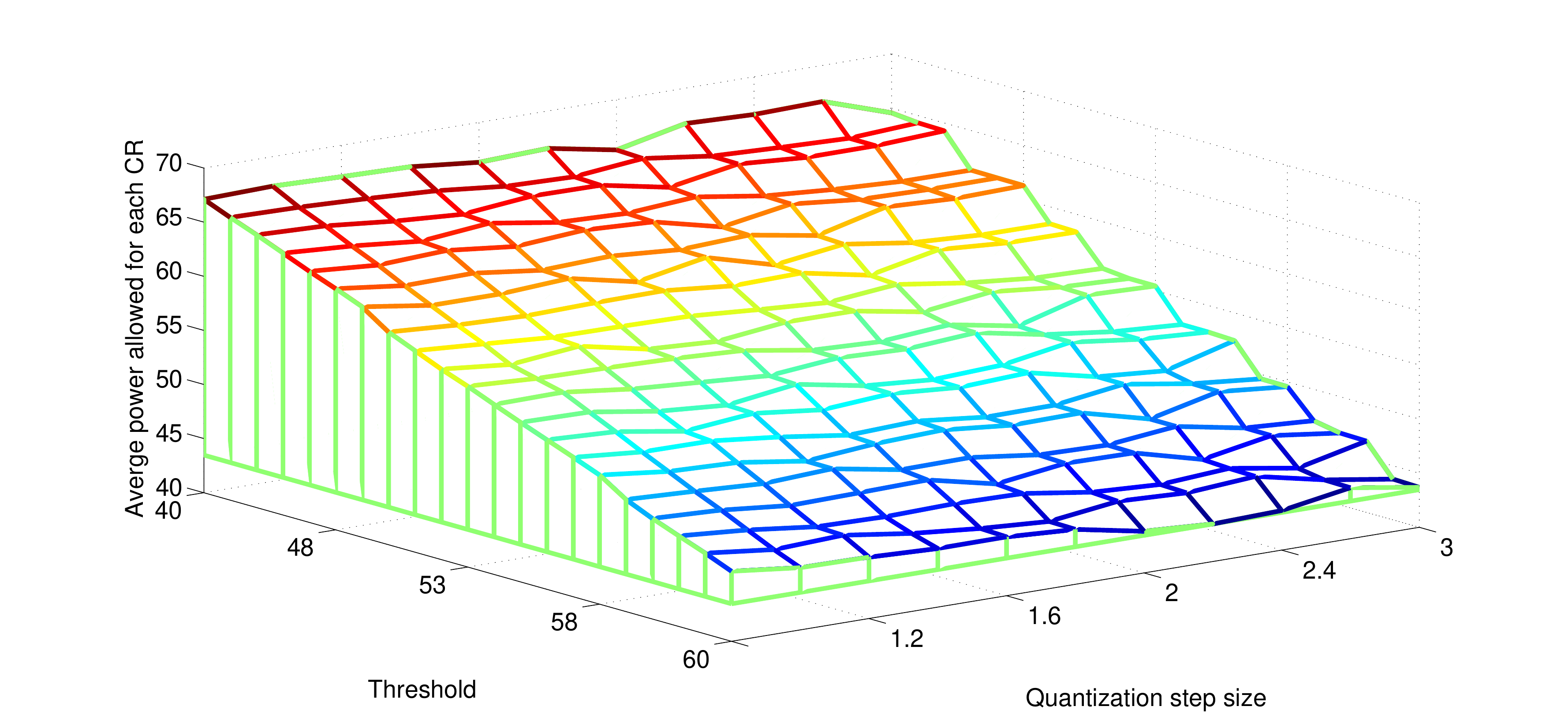}
\caption{Average allocated power for each CR vs. power quantization level and tolerable interference threshold}
\label{fig:np}
\end{figure}

Simulations were performed for 100 runs of Monte Carlo. Total number of CRs varies from 7 to 30. Power, threshold and quantization step size are in mW. Maximum power budget of each CR is set to 100 mW. In Figures~\ref{fig:pow1},~\ref{fig:cycle1},~\ref{fig:msg1} power quantization step size is set to 2 mW, and equal interference threshold set by CRs varies. Locations of CRs relative to each other, which determine the channel response gains are generated randomly. Furthermore, interference thresholds of CRs are considered to have uniformly random distributions. Figure~\ref{fig:pow1} shows average maximum allocated power to each CR after the solution is achieved. As expected, higher interference thresholds allow CRs to use more of their power budget for transmission. Also, as the number of CRs in the network increases, due to increased number of constraints, maximum allowed power for each CR decreases. Total number of cycles required to solve the distributed constraint satisfaction problem among CRs is shown in Figure~\ref{fig:cycle1}. One cycle consists of reading all incoming messages, testing local constraints, and then sending messages. Obviously, larger numbers of CRs require more cycles for algorithm run. Also, since lower interference thresholds imply more constrained transmission, they increase total cycles to reach a solution. Figure~\ref{fig:msg1} contains exchanged messages for CRs vs. interference threshold and total number of CRs. Note that algorithm run time is much less than number of exchanged messages, on account of the fact that CRs can exchange multiple concurrent messages in one cycle. To study effects of power quantization step size on number of messages passed among 7 CRs consider Figure~\ref{fig:quant}. Clearly, increasing the step size decreases total number of messages. Figure~\ref{fig:np} depicts allocated transmission powers for CRs when the number of quantization step sizes varies from 1 to 3 mW. Results indicate the protocol converges very fast to optimal solution, which makes it suitable for practical applications.


\section{Conclusion}
\label{sec:conclus}

In this paper a lightweight decentralized protocol for robust scheduling and power control in \textit{ad-hoc}  cognitive STDMA and CDMA networks was developed. CRs need not know channel gains in this method and the algorithm is based on exchanging local messages. In STDMA networks our scheme has two stages. In our multi stage protocol, as a first step, CRs interact with primary, in an iterative manner, to make sure they do not impose interference in underlay PU and SU coexistence. Then, CRs use AWCS among themselves to reach optimal resource allocations. 
For cognitive CDMA networks, the two cases of equal and unequal rates were investigated. When equal rates can be assigned to CRs, it was shown how AWCS can be used for optimal power allocation meeting QoS needs of nodes, considering their constraints on each other. AWCS with multiple local variables was applied to assign optimal rate and powers to CR nodes, free of a central management.  

For cognitive STDMA networks, where multiple CRs can use same time slots and frequencies, provided they are spatially apart to avoid interference, our protocol includes a third step for scheduling, in addition to the first and second steps envisioned for cognitive CDMA networks. To this end, the protocol identifies disjoint sets of interfering CRs with their CR head and uses inter-frame circular round robin ordering.

Simulation results corroborate benefits of this protocol for practical applications, in terms of reasonable algorithm cycles, meeting QoS constraints of nodes, and overcoming the need to know mutual interference channel gains, in situations where centralized management is not functional, for various reasons.

This method of autonomous network management, with some modifications, can be applied to LTE SONs and also capillary networks that realize Internet of Things.

\bibliography{IEEEabrv,dcs}
\bibliographystyle{ieeetr}

\end{document}